\documentclass{ptptex}

\usepackage{graphicx}
\usepackage{wrapft}

\newcommand{\beq}{\begin{equation}}
\newcommand{\eeq}{\end{equation}}
\newcommand{\beqn}{\begin{eqnarray}}
\newcommand{\eeqn}{\end{eqnarray}}

\def\agt{\mathrel{\raise.3ex\hbox{$>$}\mkern-14mu\lower0.6ex\hbox{$\sim$}}}
\def\alt{\mathrel{\raise.3ex\hbox{$<$}\mkern-14mu\lower0.6ex\hbox{$\sim$}}}

\title{%        %You can use \\ for explicit line-break
Funnel-flow accretion onto highly magnetized neutron stars and shock 
generation%
}

\author{%       %Use \scshape  for the family name
Shigeyuki \textsc{Karino}$^{1,2,3}$ ,
Motoki \textsc{Kino}$^{2,4}$ 
and John C. \textsc{Miller}$^{2,5}$
}

\inst{%         %Affiliation, neglected when [addenda] or [errata]
$^{1}$JAD Program, Universiti Industri Selangor,  \\
Block 4, Jln Zirkon A7/A, Seksyen 7, Shah Alam
40000, Selangor, Malaysia \\
$^{2}$SISSA/ISAS \& INFN, via Beirut 2-4, 34014 Trieste, Italy \\
$^{3}$Shibaura Institute of Technology, Tokyo, Japan \\
$^{4}$JAXA/ISAS, 3-1-1 Yoshinodai, Sagamihara, Japan \\
$^{5}$Department of Physics (Astrophysics), University of Oxford, UK
}

\recdate{Mmmmm DD, YYYY}%            %Editorial Office will fill in this.

\abst{%         %this abstract is neglected when [addenda] or [errata]

In this paper, we initiate a new study of steady funnel-flow accretion 
onto strongly magnetized neutron stars, including a full treatment of 
shock generation. 
As a first step, we adopt a simplified model considering the flow within 
Newtonian theory and neglecting radiative pressure and cooling. 
The flow is taken to start from an accretion disc and then to follow 
magnetic field lines, forming a transonic funnel flow onto the magnetic 
poles. 
A standing shock occurs at a certain point in the flow and beyond this 
material accretes subsonically onto the star with high pressure and density.
We calculate the location of the standing shock and all other features of 
the flow within the assumptions of our model. 
Applications to observed X-ray pulsars are discussed. 

}

\begin{document}

\maketitle

% * * * * * * * * * * * * * * * * * * * * * * * * * * * *
\section{\label{sec:1}Introduction}
% * * * * * * * * * * * * * * * * * * * * * * * * * * * *

Following the discovery of periodically variable X-ray sources \cite{g71,t72} 
and subsequent theoretical \cite{pr72,lpp73} explanations, systems consisting 
of strongly magnetized neutron stars accreting from their binary companions 
have become widely accepted as providing the model for periodic X-ray 
sources.\cite{wsh83,n89,b97} The accreted matter coming via Roche lobe overflow 
from the companion forms an accretion disc around the neutron star (NS). If the 
NS has a strong magnetic field ($\agt 10^{12} \rm{G}$), eventually at least 
some of the accreting matter cannot continue to accrete directly onto the NS 
surface in the disc plane but rather is constrained to follow the magnetic 
field lines, forming a funnel flow onto the magnetic poles (e.g. 
Ref.~\citen{pr72}). The polar caps reach high temperatures, emitting 
high-energy radiation in the form of X-rays. If the magnetic axis and the 
rotational axis are misaligned, this high energy radiation from the magnetic 
poles may be observed as periodic pulsations. \cite{st83,wsh83,n89,b97}

Polar accretion models assuming strong magnetic fields have been studied by 
many authors. \cite{pr72,lpp73,i75,bs76}\cite{gpl77,t77,gl78,lr82,klur02} In 
these studies, it has been shown that accretion onto the polar regions of 
strongly magnetized NSs and shock generation near to the NS surface can explain 
the high energy emission seen in X-ray pulsars. Almost all previous studies, 
however, have been limited to cylindrical or conical funnel flow aligned with 
the magnetic axis, whereas the actual accretion flow must have a curved 
geometry along the magnetic field lines going from the disc to the stellar 
pole. Including the effects of the curved geometry may well be very important 
for understanding the complex behaviour of the light curves of X-ray pulsars. 
\cite{ww81,n81} We here solve the flow equations in a consistent way, starting 
from where the funnel-flow material leaves the plane of the accretion disc and 
following it until it impacts on the magnetic poles.

There have been several previous studies of this type which have included the 
effects of the curved flow geometry. \cite{l96,pc96,lw99,klur02} All of these 
solve the Bernoulli equation, obtaining stationary transonic flow solutions 
similar to Bondi accretion. \cite{b52} In the present study, we mainly follow 
the approach of Koldoba et al. (2002, hereafter, KLUR). \cite{klur02} 
(We note that Koldoba and collaborators have subsequently carried out very 
interesting related calculations using an MHD computer code \cite{rukl02} but 
these were not 
addressing the question of shock location which is our main interest here.) 
Assuming a dipole magnetic field whose axis coincides with the stellar rotation 
axis, KLUR constructed the equations for transonic flow and calculated
accretion flows which accelerate from the subsonic regime into a 
supersonic regime. In their study, however, although they found the location of 
the transonic critical point, they did not calculate the position of the shock 
front. In the present work, we now extend this, following the same basic 
picture but calculating also the shock location and strength and the effects of 
shock heating. This represents the first step in a larger project to study this 
mechanism in full detail.

We adopt here a simplified model, considering a one-dimensional steady funnel 
flow along the field lines of a strictly dipole magnetic field aligned with the 
rotation axis, treating it within Newtonian theory and neglecting radiative 
pressure and cooling. Since the field is taken to be a purely dipole one (in 
the region of the funnel flow) we do not have any Poynting flux term in our 
calculation. We require the field to be strong enough to keep the funnel 
``rigid'' and to keep the fluid flowing along the field lines: under the 
circumstances envisaged, this condition should be well-satisfied everywhere 
apart from close to where the flow leaves the disc and where it impacts on the 
NS surface. 

We recognize that the simplifications which we are making are very significant 
ones but they give us a problem which can then be solved completely in a 
straightforward way, providing a useful basis for future work. Within the 
picture which we are using here, we can calculate the properties of the flow 
all the way from where it leaves the plane of the disc up to the stellar 
surface, allowing us to see several aspects which have not been observed in 
previous studies. In particular, we have shown that the shock location is 
sensitively dependent on the ratio of the pressures at the points where the 
flow leaves the disc and where it impacts on the NS surface. This is a crucial 
property of the accreting systems. Determining the shock location and the 
global characteristics of the flow is important for linking the properties of 
observed X-ray pulsars to the physical conditions in the NS -- disc systems.

We want to stress that our approach here does not attempt to include any detailed 
treatment of the physics near to the neutron star surface or near to where the 
funnel flow moves out of the plane of the disc. 
This has been treated elsewhere 
by a number of authors (as described in a later section). 
We focus instead on 
how the global solution fits together, something which we consider as certainly 
forming a necessary framework within which to fit these other more detailed 
discussions of particular parts of the flow.

The plan of this paper is as follows: in Section~\ref{sec:2} we describe how we 
set up the problem. Section~\ref{sec:3} then introduces the formulation for our 
treatment of transonic flow along the dipole magnetic field lines and 
Section~\ref{sec:4} presents the equations for treating the standing shock. 
Section~\ref{sec:5} describes our method of solution for flows including shocks 
and presents our results. In Section~\ref{sec:6}, several related issues are 
discussed. Finally, Section~\ref{sec:7} contains a summary of the 
paper together with some further discussion of the results and their 
astronomical implications. Throughout the paper, we use standard spherical 
coordinates $(r, \theta, \varphi )$ as well as cylindrical coordinates $(R, 
\varphi, z)$, which are convenient for some of the descriptions. The subscripts 
``NS'' and ``d'' denote, respectively, that the values of the physical 
parameters concerned are measured at the NS surface or at the point where the 
funnel flow leaves the plane of the disc.

% * * * * * * * * * * * * * * * * * * * * * * * * * * * *
\section{Problem set-up}\label{sec:2}
% * * * * * * * * * * * * * * * * * * * * * * * * * * * *

We are considering here a strongly magnetized neutron star with a dipole field 
surrounded by an accretion disc. At some point, moving in through the disc, the 
accretion flow becomes significantly altered due to its interaction with the 
magnetic field which, in turn, is distorted away from its purely poloidal form 
within the disc because of interaction with the disc material (see Kluzniak \& 
Rappaport \cite{kr07} for a recent treatment). Eventually, at least some part 
of the accreting material starts to go along the dipole field lines onto the 
magnetic poles of the star. As with the previous studies, we consider here a 
{\it rigidly rotating dipole field} everywhere within the region of the funnel 
flow, co-rotating with the neutron star. \cite{pc96,l96,lw99,klur02} For a 
sufficiently strong magnetic field, each element of the accreting matter will 
fall onto the star along a single field line passing through its initial 
position on the disc. The geometry of the dipole field can be described by
 \beq
r = R_{\rm{d}} \sin^2 \theta ,
\eeq 
 where, $R_{\rm{d}}$ is the distance from the centre of the star to the point 
where the funnel flow leaves the plane of the disc.

The dipole field strength can be written as  
\beq
B_{\rm{p}}^2 = \frac{\mu^2}{r^6} 
\left( 4 - \frac{3 r}{R_{\rm{d}}} \right) ,
\label{eq:bprd}
\eeq 
 where $B_{\rm{p}}$ is the poloidal magnetic field and $\mu$ is the dipole 
moment. This equation gives the magnetic field strength as a function of radial 
coordinate $r$ along a field line which intersects the disc at $r = 
R_{\rm{d}}$. The location of the point at which the funnel flow leaves the disc 
is determined by the competition between pressure, gravity and magnetic forces 
and is hard to estimate precisely, particularly in view of instabilities which 
play a role in this. We therefore simply choose suitable values for 
$R_{\rm{d}}$ and carry out our calculations with these (see Section 5).

Following KLUR, we note that transonic solutions are only possible for 
$R_{\rm{d}}$ close to the co-rotation radius $R_{\rm{corot}}$ (i.e., where the 
disc is co-rotating with the neutron star) and so we make the approximation of 
assuming that the stellar angular velocity $\Omega$ coincides with the 
Keplerian rotation rate at $r = R_{\rm{d}}$; i.e.,
 \beq
\Omega^2 = \frac{G m}{R_{\rm{d}}^3} , 
\label{eq:om}
\eeq 
where $G$ is the gravitational constant, and $m$ is the mass of the star.

Under these conditions, the transonic accretion flow can be solved for 
using the method given by KLUR (see Section~\ref{sec:3}). This flow will 
have a shock, at a certain point in the supersonic region, which we here 
assume to be adiabatic and planar. Next, we introduce the further 
assumption that there are {\it no radiative energy losses} from the 
inflowing material; the Bernoulli equation can then be applied 
everywhere along the flow-lines.

Following these preparations, we can then solve for the whole stream 
configuration, including an adiabatic shock, if we impose suitable boundary 
conditions at the point where the funnel flow leaves the plane of the disc and 
at the stellar surface. 
The main goal of the calculation is to estimate the position of 
the standing shock and the distribution of physical quantities in 
the flow.

% * * * * * * * * * * * * * * * * * * * * * * * * * * * *
\section{Formalism for flow without a shock}\label{sec:3}
% * * * * * * * * * * * * * * * * * * * * * * * * * * * *

Following KLUR, we solve the energy conservation equation (Bernoulli equation) 
with the condition that each fluid element moves along a magnetic field line 
going from $r = R_{\rm{d}}$ on the disc to the NS surface.  For the present 
situation, we write the Bernoulli equation (in the reference frame co-rotating 
with the field line) as
 \beq
E = \frac{1}{2} v^2 + \frac{a^2}{\gamma - 1} - \frac{G m}{r} 
- \frac{1}{2} \Omega^2 r^2 \sin^2 \theta ,
\label{eq:3}
\eeq 
 where $E$ is the specific energy (assumed to be constant), $v$ is the poloidal 
velocity along the field line, $a$ is the sound speed and $\gamma$ is the 
adiabatic index. The last term in this plays the role of a potential 
corresponding to the centrifugal force. Since we are considering a {\it rigidly 
rotating dipole field} throughout the region of the funnel flow there is, by 
assumption, no toroidal field component present there. This assumption (which 
is a rather reasonable one for most of the funnel flow) allows us to omit the 
Poynting flux term which would otherwise appear in Eq.~(\ref{eq:3}). The role 
of the magnetic field then consists only in fixing the configuration of the 
flux tube, in keeping the fluid elements moving on it and in determining the 
variation with position of $\rho v$ via the MHD mass conservation equation
 \beq
4 \pi \rho \frac{v}{B_{\rm{p}}} = K ,
\label{eq:mhd}
\eeq 
(where $K$ is a constant). \cite{lmms86}
By using Eqs.~(\ref{eq:bprd}) and 
(\ref{eq:mhd}), the kinetic energy of the fluid related to motion 
along the field line can be rewritten as 
$v^2 /2 = K^2 B_{\rm{p}}^2 / 32 \pi^2 \rho^2$ and inserting this into 
the Bernoulli equation gives 
\beqn
E = \varepsilon (r, \rho) =
\frac{K^2 \mu^2}{32 \pi^2 \rho^2 r^6} 
\left( 4 - \frac{3 r}{R_d} \right)  %%^{1/2} 
+  \frac{s \rho^{\gamma - 1}}{\gamma - 1}   \nonumber \\
- \frac{G m}{r}
- \frac{1}{2} \Omega^2 r^2 \sin^2 \theta .
\label{eq:ene}
\eeqn
 Here we have used the specific entropy, $s \equiv a^2 / \rho^{\gamma - 1}$ 
instead of the sound speed. With the assumption that the specific energy $E$ 
and the specific entropy $s$ are conserved along the flow-line, this equation 
gives a relationship between $\rho$ and $r$ and so we can calculate the density 
distribution $\rho(r)$ along the flow-line if we specify boundary conditions at 
the point of departure from the disc: $E = E(R_{\rm{d}}) = \rm{const.}$ and $s 
= s(R_{\rm{d}}) = \rm{const.}$

For convenience of calculation, we rewrite Eq.~(\ref{eq:ene}) in the 
dimensionless form: 
\beq
\tilde{E} = \tilde{\varepsilon} (\tilde{r}, \tilde{\rho}) = 
\frac{4 - 3 \tilde{r}}{2 \tilde{\rho}^2 \tilde{r}^6} 
+ \frac{\tilde{s} \tilde{\rho}^{\gamma - 1}}{\gamma - 1}
- \left( \frac{1}{\tilde{r}} + \frac{\tilde{r}^3}{2} \right) . 
\label{eq:k15}
\eeq
Here, we have used following transformation of variables,
\beqn
r &=& R_{\rm{d}} \tilde{r}, \quad 
\rho = \frac{K \mu}{4 \pi \Omega R_{\rm{d}}^4} \tilde{\rho}, 
\nonumber \\
s &=& \Omega^2 R_{\rm{d}}^2 
\left( \frac{4 \pi \Omega R_{\rm{d}}^4}{K \mu} \right)^{\gamma - 1} 
\tilde{s}, \quad 
E = \Omega^2 R_{\rm{d}}^2 \tilde{E} .
\label{eq:norm}
\eeqn 
 (From here on, since we use only dimensionless quantities, we will omit the 
tildes.) By solving Eq.~(\ref{eq:k15}), we can obtain the values of the 
physical quantities everywhere along a flow-line if there is no shock. Note 
that when working in terms of the dimensionless quantities, the field strength 
does not itself enter the calculation, only its variation with position.

We are interested in transonic solutions with the initially 
subsonic inflow becoming supersonic at a critical point. In order that 
the flow should be non-singular at the critical point (as required for a 
physically meaningful steady-flow solution) regularity conditions must 
be satisfied there, as given by Eqs. (16) and (17) of KLUR.

% * * * * * * * * * * * * * * * * * * * * * * * * * * * *
\section{The treatment of shocks}\label{sec:4} 
% * * * * * * * * * * * * * * * * * * * * * * * * * * * *

We will denote the location of the critical point by $r_{\rm{c}}$; a 
shock may then appear beyond this in the supersonic part of the flow 
with the location depending on the boundary conditions. 
We next summarize our methodology for treating the shock within our 
simplifying assumptions of considering a stationary planar shock which 
is adiabatic (i.e. no energy is lost from the flow). \cite{ll59}
From the conservation of the mass, momentum and energy fluxes across 
the shock, one obtains the Rankine-Hugoniot relations linking the values 
of quantities on either side of it. 
The ones of relevance for us here are:
\beqn
\rho_2 &=& \frac{(\gamma + 1) M_1^2}{(\gamma - 1) M_1^2 + 2} \rho_1 
\label{eq:rho2} \\
p_2 &=& \frac{2 \gamma M_1^2 - (\gamma - 1)}{\gamma + 1} p_1 
\label{eq:pre2} 
\eeqn 
where $M=v/a$ is the Mach number and the subscripts 1 and 2 denote the 
values of flow quantities immediately before and after the shock. 
The shock raises the specific entropy of the fluid, with the change 
being given by
\beq
\Delta s = s_2 - s_1  
= \gamma \left( \frac{p_2}{\rho_{2}^{\gamma}} 
- \frac{p_1}{\rho_{1}^{\gamma}} \right) .
\label{eq:s-en}
\eeq 
In the next section, we use these shock relations together with the 
flow equation (\ref{eq:k15}).

% * * * * * * * * * * * * * * * * * * * * * * * * * * * *
\section{Accretion flow with a shock}\label{sec:5}
% * * * * * * * * * * * * * * * * * * * * * * * * * * * *

\subsection{Flow beyond the shock}\label{sec:51}

The behaviour of the fluid quantities along the flow-line, for given boundary 
conditions at $r = R_{\rm{d}}$, is completely 
described by the Bernoulli equation (\ref{eq:k15}) except that, when 
passing across the shock front, the value of the specific entropy needs 
to be changed to the new higher value, i.e. beyond the shock, the 
Bernoulli equation becomes:
\beq
E = \varepsilon (r, \rho) =
\frac{4 - 3r}{2 \rho^2 r^6} + \frac{s_{2} \rho^{\gamma - 1}}{\gamma - 1}
- \left( \frac{1}{r} + \frac{r^3}{2} \right) , \label{eq:k15b}
\eeq 
where $s_{2}$ is the specific entropy of the flow beyond the shock, 
given by Eq.~(\ref{eq:s-en}).

Our procedure for calculating the entire flow solution with the shock, starting 
from conditions at $r = R_{\rm{d}}$ on the disc and ending at the surface of 
the NS, is as follows:
(i)~Choose a position for the shock $r = r_{\rm{s}}$, with  
$r_{\rm{NS}} \le r_{\rm{s}} \le r_{\rm{c}}$. 
(ii)~Solve for the flow in the region upstream of the shock, 
$r_{\rm{s}} \le r \le R_{\rm{d}}$, using Eq.~(\ref{eq:k15}). 
This gives the value of $\rho_1 = \rho(r_{\rm{s}})$, just before 
the shock, from which can be calculated the corresponding values 
of $p_1$ and $M_1$. 
(iii)~Using the Rankine-Hugoniot relations, we then calculate the 
values of density and pressure beyond the shock ($\rho_2$ and $p_2$) and 
hence the new value of the specific entropy $s_2$. 
(iv)~Using this value of $s_{2}$, we then calculate the flow 
beyond the shock using Eq.~(\ref{eq:k15b}).
In this procedure we obtain the value of $\rho$ at the stellar 
surface and hence also the values of the 
other fluid parameters there. 
(v)~In practice, we would prefer to specify the value of the 
pressure at the stellar surface rather than the location of the shock 
and so we iterate the solution until the shock location corresponding to 
the desired surface pressure is found.

\begin{figure}
\begin{center}
\includegraphics[width=8cm]{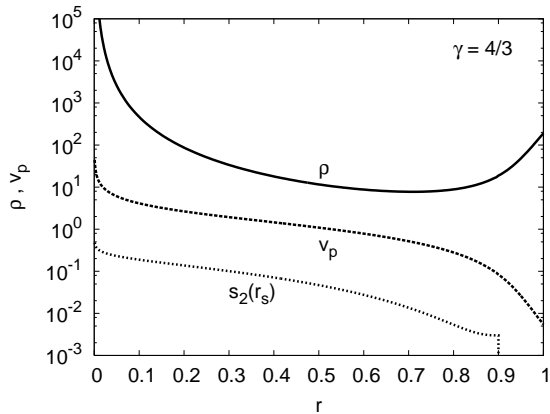} 
\caption{Density and poloidal velocity distributions along 
flow-lines with no shock for different values of $\gamma$. 
We show the case with $\gamma = 4/3$. 
Behaviour of the the enhanced entropy which would appear 
beyond shocks occurring at different locations $r_{\rm{s}}$ 
is also shown.
\label{fig:newfig1} }
\end{center}\end{figure}

In Fig.~\ref{fig:newfig1}, we show our solutions for density and velocity in
the special case of transonic flow with {\it no} shock. 
We have made calculations for $\gamma = 4/3$ with parameter 
values of $s = 0.003$ and $E = -1.448$.
Note that $\gamma = 4/3$ is smaller than the maximum value of $\gamma$ 
for which transonic flow can occur in Bondi accretion ($\gamma = 5/3$).
This figure also shows the increased entropy $s_2$ which would appear 
beyond shocks occurring at different locations $r_{\rm{s}}$ along the flow-lines.
The value of $s_2$ is one of the most important quantities here, 
since it controls the whole flow solution in the downstream region, 
$r_{\rm{NS}} \le r \le r_{\rm{s}}$. 

\begin{figure}
\begin{center}
\includegraphics[width=8cm]{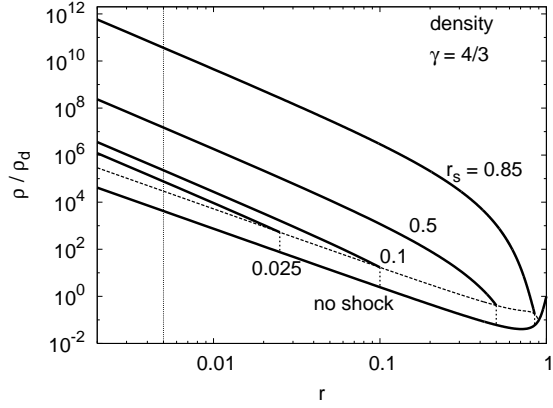}
\caption{Density distributions for flows including shocks, taking 
$\gamma = 4/3$ and $s = 0.003$. 
The upper solid curves are for flows with shock discontinuities at 
$r = 0.85, 0.5, 0.1$ and 0.025 (with the dashed lines marking the shock 
discontinuities); the dotted curve is the envelope of the density 
values on the downstream side of the shock; 
the lowest thick solid curve is for a flow with no shock. 
The vertical thin-dotted line denotes the position of the NS surface
in the case of $r_{\rm{NS}} / R_{\rm{d}} = 5.0 \times 10^{-3}$.
\label{fig:newrho} }
\end{center}
\end{figure}

\begin{figure}
\begin{center}\includegraphics[width=8cm]{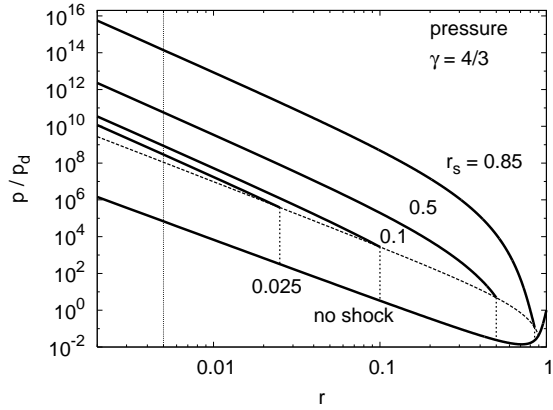}
\caption{Pressure distributions corresponding to the density 
distributions in Fig.~\ref{fig:newrho}. 
\label{fig:newpre} }
\end{center}\end{figure}

Figs.~\ref{fig:newrho} and \ref{fig:newpre} show 
the variation of density and pressure along the flow-lines; the lowest 
solid curve corresponds to transonic flow without a shock, as shown in 
Fig.~\ref{fig:newfig1}, and the upper curves are for flows having an 
adiabatic shock at different locations. When there is a shock, density 
and temperature have discontinuous jumps across it, as given by 
Eqs.~(\ref{eq:rho2}) and (\ref{eq:pre2}). The envelope of the density 
values on the downstream side of the shock, $\rho_2$, is shown by the 
dotted curve in Fig.~\ref{fig:newrho}, and the envelope for $p_2$ 
is shown in the same way in Fig.~\ref{fig:newpre}. The behaviour of 
the physical quantities for $r_{\rm{NS}} \le r \le r_{\rm{s}}$ is given 
by Eq.~(\ref{eq:k15b}). Note that the density can increase across the 
shock by at most a factor of $(\gamma +1)/(\gamma - 1)$, within 
the non-relativistic theory being used here, whereas increases of 
several orders of magnitude may occur by means of progressive 
compression in the subsonic flow beyond the shock.

From Figs.~\ref{fig:newrho} and \ref{fig:newpre}, we see that the resulting 
density and pressure at the stellar surface depend sensitively on the location 
of the shock (as an example, $r_{\rm{NS}} = 0.005 R_{\rm{d}}$ is shown as the 
vertical dotted line in the figures). As mentioned previously, if the pressure 
(or density) at the stellar surface is given as a boundary condition, the shock 
location can then be determined and the whole solution completed.
 Note that in Figs.~\ref{fig:newrho} and \ref{fig:newpre}, the curves for the 
higher values of $r_{\rm{s}}$ rise quickly just beyond the shock and then 
flatten to an almost constant slope, whereas those for smaller $r_{\rm{s}}$ 
have an almost constant slope throughout the region beyond the shock. The 
reason for this difference can be understood from Eq.~(\ref{eq:k15b}): in 
calculating the behaviour of the density and pressure just beyond the shock, 
the gravity term $ \propto r^{-1} $ completely dominates over the centrifugal 
term $ \propto r^{3}$ when the shock occurs near to the stellar surface, 
whereas the centrifugal term makes a significant contribution with respect to 
the gravity term further from the star. This difference is reflected in the 
two-component gradients of the curves in Fig.~\ref{fig:new}, introduced below.

\subsection{Pressures and densities achievable at the NS surface}

Before going into details about the shock location, we here discuss
the range of possible densities in the flow at the point where it impacts 
on the NS surface. 
The {\it maximum} achievable density corresponds to the shock 
occurring just after the flow passes the critical point. 
In this case, the fluid reaches the NS surface with almost zero poloidal 
velocity and so the kinetic energy term in Eq.~(\ref{eq:k15b}) 
can be neglected there\footnote{ This can be easily understood if 
we consider the analogy with Bondi accretion. 
Bondi flow has two trans-sonic sequences, one representing flow which 
is being {\it accelerated} from sub-sonic to super-sonic and the other 
representing flow being {\it decelerated} from super-sonic to sub-sonic. 
An accretion-flow solution starts off along the first (accelerated) 
sequence but, after the critical point, a shock may occur in which 
case the solution jumps down to the second (decelerated) sequence, 
where it is then sub-sonic. 
Following this, the flow is progressively decelerated. If the shock 
occurs very close to the critical point, there is a relatively long 
deceleration time before the flow reaches the stellar surface and so 
the velocity there will be quite low. }. 
The rotational kinetic energy term at the NS surface can also be neglected 
since the field lines are taken to be  
co-rotating with the neutron star which is itself slowly-rotating
for all of the cases of interest here. We then obtain 
\beq 
\rho_{\rm{NS,max}} 
\approx \left[ \frac{\gamma - 1}{s_2} 
\left( E + \frac{R_{\rm{d}}}{r_{\rm{NS}}} \right) \right]^{1/(\gamma - 1)}. 
\label{eq:analy1} 
\eeq 
The {\it minimum} achievable density corresponds to the shock occurring 
just above the stellar surface. In this case, the kinetic energy 
dominates over the thermal energy just before the shock and we obtain 
\beq
\rho_{\rm{NS,min}} \approx 
\frac{(\gamma +1) M_{1, \rm{NS}}^2}{(\gamma - 1) M_{1, \rm{NS}}^2 +2} 
\times \left[ 
\frac{R_{\rm{d}}^5 (4 R_{\rm{d}} - 3 r_{\rm{NS}}) } 
{2 r_{\rm{NS}}^5 (E r_{\rm{NS}} + R_{\rm{d}}) } \right]^{1/2} . 
\label{eq:analy2} 
\eeq 
The first term on the right hand side is the factor appearing in 
Eq.~(\ref{eq:rho2}).

For given outer boundary conditions ($E, s_1, R_{\rm{d}}$), the density 
in the flow at the NS surface must lie within the range between these 
maximum and minimum values for any flow including a shock. 
The range of possible pressures in the flow at the NS surface can be 
obtained in the same way.

\subsection{Location of the shock}

Figs.~\ref{fig:newrho} and \ref{fig:newpre} show values for the 
normalized density and pressure along the flow-lines, plotted against the 
normalized radial coordinate %%%$\tilde{r} \equiv r/R_{\rm d}$ 
for various shock locations. 
We take a canonical value of 10 km for the neutron star 
radius but this then corresponds to different values of $\tilde{r}$ 
depending on the value taken for $R_{\rm d}$, the location at which the funnel 
flow leaves the disc. (For our calculations, we 
use the following representative values for the other neutron star parameters: 
$m = 1.4 M_{\odot}$, $\mu = 10^{30} \rm{G ~cm}^{-3}$ giving 
$B_{\rm{p,NS}} \approx 2.0 \times 10^{12} \rm{G}$ - see, for example, 
Shapiro and Teukolsky 1983). \cite{st83}

\begin{figure}
\begin{center}
\includegraphics[width=10cm]{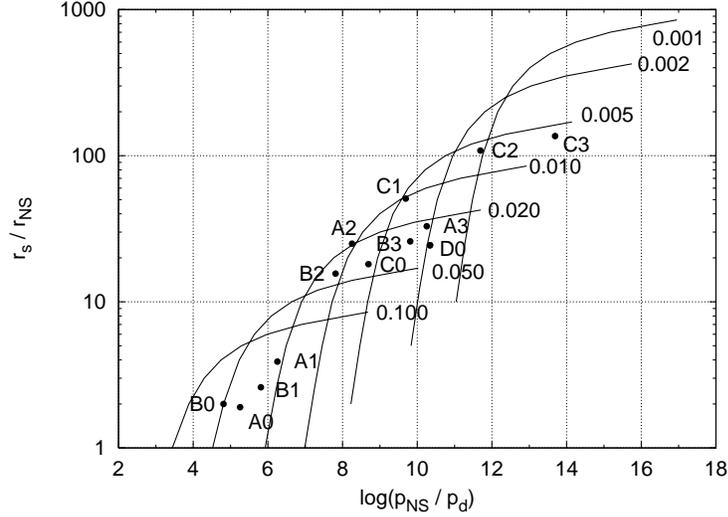} 
\caption{
Shock locations as functions of the pressure ratio. The filled circles 
correspond to the cases listed in Table~\ref{tab:1} as indicated by the 
adjacent labels. The trend for the shock moving further out as the pressure 
ratio increases was to be expected since the stopping power of the back 
pressure becomes greater as the pressure ratio increases, causing it to be felt 
further out. The reason for the change in slope seen within each individual 
curve has been discussed in Section~\ref{sec:51}.
\label{fig:new} }
\end{center}
\end{figure}

At the outer and inner boundaries of the funnel flow we impose 
pressure-matching boundary conditions. When we know the starting point of the 
flow ($R_{\rm{d}}$) and the ratio of the pressures at the inner and outer 
boundaries ($p_{\rm{NS}} / p_{\rm{d}} $), we can obtain the appropriate shock 
position. Fig.~\ref{fig:new} shows the shock position $r_{\rm{s}}$, normalized 
by the NS radius $r_{\rm{NS}}$, plotted against the pressure ratio $p_{\rm{NS}} 
/ p_{\rm{d}} $ for a range of selected values of $R_{\rm{d}}$ ($r_{\rm{NS}} / 
R_{\rm{d}} = 0.1, 0.05, 0.02, 0.01, 0.005, 0.002$ and 0.001, as marked next to 
the related curves), taking account of the fact that the actual values of 
$R_{\rm{d}}$ are rather uncertain. The shock positions (in a log scale) are 
distributed within a band roughly proportional to $\log(p_{\rm{NS}} / 
p_{\rm{d}})$: as the pressure ratio becomes larger, the shock moves further 
away from the NS surface. The shock positions obtained using the various 
specific boundary conditions discussed below are indicated by the filled 
circles in this figure. We now turn to discussion of these conditions.

\subsubsection{Outer boundary condition}\label{sec:5.3.1}

At the outer boundary $r = R_{\rm{d}}$, we set the pressure in the funnel flow 
equal to the gas pressure at that point in the disc. We discuss here a number 
of possible models for this although all of them should be considered as being 
very approximate.

First, we consider the picture where the outer boundary of the funnel flow is 
matched onto a Shakura-Sunyaev disc \cite{ss73} with the funnel flow being 
taken to start from the point where the magnetic pressure of the dipole field 
balances the gas pressure. \cite{pr72} This gives 
\begin{eqnarray} 
p_{\rm{d}}=\frac{B_{\rm{p}}(R_{\rm{d}})^2}{8 \pi} 
&=& 4.65 \times 10^{15} \alpha^{-9/10}\nonumber \\ 
&\times& 
\left( \frac{R_{\rm{d}}}{10^7 \rm{cm}} \right)^{-51/20} 
\left( 1 - 0.35 \sqrt{\frac{10^7 \rm{cm}}{R_{\rm{d}}}} \right)^{4/5} . 
\end{eqnarray} 
 where $\alpha$ is the alpha viscosity parameter (here we restrict attention 
to the situation for $\alpha = 0.1$). Using $B_{\rm{p}}(R_{\rm{d}})^2/8 
\pi = \mu^2/8 \pi R_{\rm{d}}^6$ and fixing $\dot{m}_{\rm{crit}}$, this gives 
the value of $R_{\rm d}$ ($\dot{m}_{\rm{crit}}$ is the critical accretion rate 
which gives the Eddington luminosity). We consider models with two values of 
$\dot{m} / \dot{m}_{\rm{crit}}$: ``Model A'' with $\dot{m} / 
\dot{m}_{\rm{crit}} = 2/3$ and ``Model B'' with $\dot{m} / \dot{m}_{\rm{crit}} 
= 1$.

Our third model (``Model C'') is based on the disc model of Kluzniak \& 
Rappaport \cite{kr07} who derived a modified form of the Shakura-Sunyaev 
solution for discs around magnetized stars, taking account of the way in which 
the magnetic field modifies the flow in the inner regions of the disc and is, 
in turn, modified by its interaction with the flow. They used two forms of 
ansatz for the toroidal component of the field produced in the disc (we follow 
their first prescription here), and calculated the effect on the flow of the 
resulting magnetic torque which progressively substitutes the standard viscous 
torque as one moves in through the disc. This gives a set of formulae 
for the disc profile and the variations of the fluid parameters through the 
disc which represent generalisations of the corresponding Shakura-Sunyaev 
formulae. The standard Keplerian disc then ends at the torque balance point 
$r = R_{\rm{torq}}$ (where the magnetic torque has grown to balance the 
accretion torque), i.e. where
\beq
- R_{\rm{torq}}^2 B_{\rm{p, torq}} B_{\varphi, \rm{torq}}
= \dot{m} \left[ \frac{d(R^2 \Omega)}{dR} \right]_{\rm{torq}} .
\label{eq:torq}
\eeq
 For the parameter values mentioned above, this comes very close to the 
co-rotation radius $R_{\rm{corot}}$ if the stellar rotation period is at the 
lower end of the range for X-ray pulsars. We again take the funnel flow to 
leave the plane of the disc at the point of balance between the gas pressure 
and the magnetic pressure of the dipole field; this balance occurs very close 
to the torque balance point. Our Model C has a Kluzniak-Rappaport disc with the 
same overall parameters as Model A above and has an NS rotation period of 1 
second for which $R_{\rm{torq}}$ is indeed very close to $R_{\rm{corot}}$ (and 
$R_{\rm{d}} \sim R_{\rm{corot}}$ as required for our overall picture).

Finally, we consider the situation when the outer accretion flow consists 
of matter coming essentially in free-fall from a companion star and the funnel 
flow starts at the Alfv\'{e}n radius $r_{\rm{A}}$ given by 
\begin{equation}
\frac{B^2_{\rm{p}}}{8 \pi} = \frac{1}{2} \rho(r_{\rm{A}}) v^2 (r_{\rm{A}}).
\end{equation} 
 We then take $R_{\rm{d}} = r_{\rm{A}}$ and we refer to this case as our 
``Model D''.

\subsubsection{Inner boundary condition}

At the inner boundary $r = r_{\rm{NS}}$, the fluid flow changes from being 
essentially radial to being transverse and the situation there is even less 
clear than at the outer boundary since we must consider a matching with the 
transverse pressure (including magnetic pressure) at a suitable height in the 
NS atmosphere.
 The flow pressure at the NS surface might 
then be as high as that given by balance with the magnetic pressure 
$p_{\rm{NS}} =  B_{\rm{p,NS}}^2 /8 \pi \sim 10^{23} \rm{dyn ~cm}^{-2}$.

Sophisticated NS atmospheric models give various different views about the NS 
surface pressure. For instance, Van Riper~\cite{v88} found that the density of 
the outermost layer of the atmosphere of a magnetized NS should be in the range 
$\sim 0.5 - 10 \rm{g ~cm}^{-3}$, for $m \approx 1.0 M_{\odot}$, $r_{\rm{NS}} 
\approx 10^{6} \rm{cm}$, and $B_{\rm{NS}} \approx 10^{12} \rm{G}$. However, the 
corresponding pressure is strongly temperature dependent and so even if we fix 
on $\rho \approx 1 \rm{g~cm}^{-3}$, the pressure can take a wide range of 
values going from $\approx 10^{16} \rm{dyn ~cm}^{-2}$ up to $\approx 10^{22} 
\rm{dyn~cm}^{-2}$ for the temperature ranging from $T \sim 10^7 \rm{K}$ to 
$\sim 10^9 \rm{K}$. The temperature at $r = r_{\rm{NS}}$ is almost independent 
of the shock position (as discussed in Section 5.4), giving 
%%$T_{\rm{NS}} = 1.7 \times 10^9 \rm{K}$ for $\alpha = 1.0$ and 
$T_{\rm{NS}} = 2.4 \times 10^9 \rm{K}$ for $\alpha = 0.1$. 
According to Van Riper~\cite{v88}, for $T_{\rm{NS}} 
\sim 10^9 \rm{K}$, the pressure of the NS surface layer is 
roughly $\sim 10^{21} - 10^{22} \rm{dyn ~cm}^{-2}$.

On the other hand, Brown and Bildsten~\cite{bb98} suggested that the accreted 
matter cannot spread out until it reaches a greater depth, since the magnetic 
pressure is always much larger than the matter pressure at the polar cap. In 
this case, our pressure boundary condition should be modified to $\sim 10^{25} 
\rm{[dyn~cm^{-2}]}$, and the height of the shock above the neutron-star surface 
would become much larger.

Another example of an atmosphere model for an accreting NS is 
the plane-parallel thin-slab model of Harding et al.~\cite{h84} 
According to this, the density in the atmosphere changes by about five 
orders of magnitude within a thin layer of width $\alt 200\ \rm{cm}$ 
with very little accompanying change in temperature.

Detailed consideration of the physics of the near-surface region is 
beyond the scope of the present paper which focuses instead on global 
properties of the flow. Because of the uncertainties involved, we consider here 
a range of values for $p_{\rm{NS}}$, bearing in mind the atmospheric pressures 
quoted in above references, and we will return to give a more complete 
treatment of the near-surface region in future work.

\bigskip

From Fig.~\ref{fig:new}, it can be seen that the shock location comes closer to 
the NS surface when either (1) the NS surface pressure is low, or (2) the 
pressure is high at the point where the funnel flow leaves the disc, i.e. when 
the ratio $p_{\rm{NS}} / p_{\rm{d}}$ is small, the shock will occur near to the 
NS surface. In the opposite limit, the shock will occur far above the NS 
surface, similar to the situation considered by Basko and Sunyaev. \cite{bs76}
%This behavior can be understood naturally; that is, if the pressure at 
%the NS boundary becomes high, the strong pressure pushes up the "roof" of 
%the shock front, and vice versa.
As discussed in Sec.~\ref{sec:add} below, the steep pressure gradient occurring 
in an optically thick radiative accretion column would effectively lead to a 
rather low NS surface pressure, bringing the shock nearer to the NS surface. 
Results for the various specific disc models, and for a relevant range of 
values of $p_{\rm NS}$, are listed in Table~\ref{tab:1} and depicted in 
Fig.~{\ref{fig:new}}. Cases A0, B0, C0 and D0 are the ones with the optically 
thick radiative columns.

\begin{table}
\centering
\begin{minipage}{80mm}  
\caption{Shock positions for selected surface conditions:
	Shakura-Sunyaev disc with moderate accretion (A), Shakura-Sunyaev disc 
        with critical accretion (B), Kluzniak-Rappaport disc with 
        moderate accretion (C) and the Free-fall model (D). The cases with 
        optically thick radiative accretion columns are indicated by a $\ast$, 
        and for those models, the inner boundary condition for the funnel 
        flow is imposed at the top of the column (see Sec.~\ref{sec:add}). 
}  
\label{tab:1}
\begin{tabular}{@{}cccc@{}}    
\hline \hline   
Disc Model & NS Surface Condition & Shock Position & Label \\
 &  $p_{\rm{NS}} \rm{[dyn/cm^2]}$     & $r_{\rm{s}} \rm{[cm]}$ &    \\ 
\hline 
Shakura-Sunyaev disc with & $1.0 \times 10^{18}$ ($\ast$) & $1.9 \times 10^{6}$  
& A0 \\
$\dot{m}/\dot{m}_{\rm{crit}} = 2/3$  & $1.0 \times 10^{19}$ & $3.9 \times 10^{6}$  & A1 \\
$R_{\rm{d}} = 4.4 \times 10^7$ [cm] & $1.0 \times 10^{21}$ & $25.0 \times 10^{6}$  & A2 \\
$p_{\rm{d}} = 5.6 \times 10^{12} \rm{[dyn/cm^2]}$ & $1.0 \times 10^{23}$ & $32.9 \times 10^{6}$ & A3 \\
\hline 
Shakura-Sunyaev disc with & $1.0 \times 10^{18}$ ($\ast$) & $2.0 \times 10^6$  
& B0 \\
$\dot{m}/\dot{m}_{\rm{crit}} = 1$ & $1.0 \times 10^{19}$ & $2.6 \times 10^6$  & B1 \\
$R_{\rm{d}} = 3.7 \times 10^7$ [cm]  & $1.0 \times 10^{21}$ & $15.6 \times 10^6$  & B2 \\
$p_{\rm{d}} = 1.5 \times 10^{13} \rm{[dyn/cm^2]}$ & $1.0 \times 10^{23}$ & $25.9 \times 10^6$  & B3 \\ 
\hline
Kluzniak-Rappaport disc  & $1.0 \times 10^{18} $ ($\ast$) & $18.1 \times 
10^6$ & C0 \\ 
$\dot{m}/\dot{m}_{\rm{crit}} = 2/3$& $1.0 \times 10^{19}$ & $50.9 \times 10^6$ & C1 \\
$R_{\rm{d}} = 1.6 \times 10^8$ [cm] & $1.0 \times 10^{21}$ & $108.4 \times 10^6$  & C2 \\
$p_{\rm{d}} = 2.0 \times 10^9 \rm{[dyn/cm^2]}$ & $1.0 \times 10^{23}$ & $136.3 \times 10^6$  & C3 \\ 
\hline
Free-fall model & $1.0 \times 10^{18}$ ($\ast$) & $22.4 \times 10^6$ & D0 \\
$R_{\rm{d}} = 3.1 \times 10^8$ [cm] \\
$p_{\rm{d}} = 4.6 \times 10^7 \rm{[dyn/cm^2]}$ \\ 
\hline
\end{tabular}
\end{minipage}
\end{table}

\section{Discussion of related issues{\label{sec:6}}}

\subsection{Departure of the funnel flow from the plane of the disc}

In discussing the disc models above, we estimated the location of the point 
where the funnel flow leaves the plane of the disc by assuming that the 
magnetic pressure of the dipole field of the NS balances the pressure of the 
accreting gas there, following Pringle \& Rees \cite{pr72}. This seems a 
reasonable approximation but there are a number of issues that make the 
situation less clear. First, we should stress that there is a distinction to be 
made between the place where material starts to leave the plane of the disc and 
the place where one should define the ``inner edge'' of the disc to be, 
although they are often treated as being the same. In an accreting system, the 
``inner edge'' is not a completely clear concept although one can take it to be 
the place where the disc ceases to be nearly Keplerian in the case of a 
geometrically thin disc. Also, while one often takes the disc to be ``thin'', 
nevertheless it does have a vertical structure and conditions are certainly not 
constant through the height of it. It is likely that there will be a separation 
between some material which goes into the funnel flow while other material 
continues in or near the disc plane. Another point is that the pressure-balance 
argument is based on supposing that the poloidal field of the neutron star is 
not affected by its interaction with the disc material whereas, in fact, there 
will be at least some region where the material flow distorts the field lines 
producing a toroidal component within the disc and giving a torque on the 
accreting material, as mentioned earlier. Finally, there is the issue of a 
variety of instability mechanisms which can influence the situation. It is 
therefore clear that the question of how and where material leaves the disc 
plane to go into the funnel flow is a complicated one.

Following the pioneering study by Ghosh and Lamb,~\cite{gl78} many authors have 
worked to improve understanding of the magneto-hydrodynamic properties of 
accretion discs but this still remains a difficult subject. Frequently, the 
position of the ``inner edge'' of the disc is taken to come at the point of 
balance between the magnetic torque and the accretion torque (see the 
discussion in Section~\ref{sec:5.3.1}) with the flow remaining essentially 
Keplerian down to that point. This is the case for the Kluzniak-Rappaport model 
which we have used~\cite{kr07} and there one finds that the pressure balance 
point is extremely close to the torque balance point for parameter values of 
interest here. However, this depends on the ansatz made there for describing 
the toroidal field $B_{\varphi}$ and, while that seems a rather reasonable one, 
this is something which should, in the end, come from a calculation including 
all of the physical features concerned.
 In basic MHD treatments of discs, the effective inner radius is almost 
coincident with the co-rotation radius, $R_{\rm{corot}}$, \cite{w87,w95} but 
some more detailed studies suggest that it may be significantly inward of this, 
at $R_{\rm{torq}} \sim 0.1 R_{\rm{corot}}$. \cite{lw99} Alternatively, its 
location may be related to the Alfv\'{e}n radius, as in Model D in our 
treatment.~\cite{w96} 
%Additionally, some numerical MHD simulations suggest that the inner edge will
%be located at the balance point between the magnetic pressure and matter
%pressure.~\cite{rukl02}

Even if the gas pressure does balance the magnetic pressure at the starting 
point of the flow, the magnetic pressure would then grow very rapidly along the 
flowlines (as $\sim r^{-6}$) and would soon become dominant so as to establish 
the funnel-like geometry of the flow. This would then be expected to persist 
until the gas pressure could catch up with the magnetic pressure again after 
the flow penetrates into the NS surface layers.

\subsection{Radiative effects \label{sec:add}}

In the above treatment, we have not included the effects of radiation 
emitted by the fluid comprising the funnel flow but, in realistic 
situations, radiation pressure and radiative energy losses will be 
important near to the stellar surface. We will include these effects 
consistently in our future work but here we make just a rough estimate 
of how radiation pressure would change the present results. 
For doing this, we join our flow solution onto the solution for an 
accretion column calculated by Inoue~\cite{i75}, which does include radiation 
pressure. 
Inoue (1975) solved the fluid equations, including photon 
diffusion, for a short accretion column just above the NS surface and 
found that the density distribution can be written as 
\beq 
\rho = \rho_{\rm{NS}} \left( \frac{r}{r_{\rm{NS}}} \right)^{-6} 
\exp \left[ \frac{ 3 (r_{\rm{NS}} - r)}{r_{\rm{diff}}} \right] , 
\label{eq:inoue} 
\eeq 
where, $r_{\rm{diff}} = \kappa_{\rm{th}} \dot{m} / 8 \pi c$ with 
$\kappa_{\rm{th}}$ being the Thomson scattering opacity. This density 
distribution (and the corresponding pressure distribution) has quite a 
strong dependence on $r$ and the accretion column becomes optically 
thin rather close to the NS surface. In this solution, the pressure of 
the column decreases by roughly five orders of magnitude within the 
optically thick regime.

According to Inoue, the height of the optically thick 
radiative zone will be $\sim 6 \times 10^5 \rm{cm}$ for 
$\dot{m} = 10^{17} \rm{g~s}^{-1}$. \cite{i75}
When we join our solution at the top of the column with the density 
and pressure obtained from Eq.~(\ref{eq:inoue}), the resulting 
shock position is closer to the stellar surface. 
Taking the pressure at the top of the radiative column ($\tau = 1$) 
to be $\sim 10^{18}{\rm dyn/cm^2}$ and the outer boundary condition 
to be that for Shakura-Sunyaev disc model with 
$\dot{m} = \frac{2}{3} \dot{m}_{\rm{crit}} \simeq 10^{17} {\rm g~cm^{-1}}$ 
(Model A), the shock position is then just above the top of the radiative 
column.  
This means that the flow becomes optically thick just after the shock 
and then emits high energy photons as black body radiation. 
If, on the other hand, we take the same pressure at the top of the  
radiative column but use the free-fall condition (Model D) for the inner  
boundary condition, the shock position is then still far from 
the NS surface and from the top of the radiative column
($r_{\rm s} = 22.4 \times 10^6 {\rm cm}$). 
Hence, in this case the flow becomes dense and achieves  
high temperatures beyond the shock, but it is still optically  
thin there and so emits energy via free-free radiation or cyclotron  
processes before forming the optically-thick radiative column. 
This optically-thin, but actively radiating region corresponds to a
``transition region''. \cite{i75,bs76}

\subsection{Temperature distribution}\label{sec:temperature}

\begin{figure}
\begin{center}
\includegraphics[width=8cm]{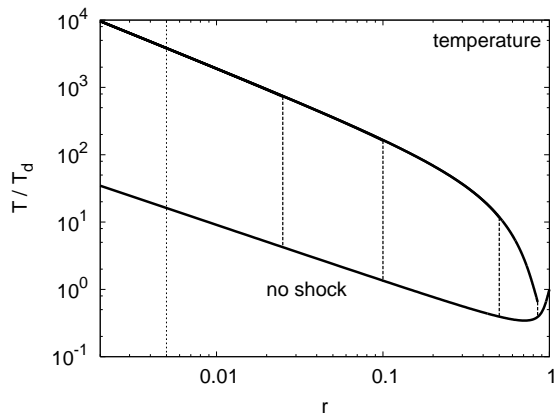}
\caption{Temperature distribution along the flow-lines, taking 
$\gamma = 4/3$. 
\label{fig:temp} }
\end{center}
\end{figure}

In this subsection, we present some results for the temperature distribution 
beyond the shock, as estimated using 
$T \equiv s \gamma^{-1} \rho^{\gamma - 1}$. 
In Fig.~\ref{fig:temp}, the temperature distribution for a transonic flow with 
no shock is shown by the lower solid curve and the corresponding temperature 
distributions for shocked flows with the selected values of $r_{\rm{s}}$ are 
also shown (cf Figs.~\ref{fig:newrho} and \ref{fig:newpre}). It is interesting 
that the temperature curves beyond the shock are independent of the shock 
position. This can be understood from the solution of the Bernoulli equation: 
since the kinetic energy is negligible beyond the shock, we have
\beq 
T \propto \frac{\gamma - 1}{\gamma} 
\left[ E + \frac{1}{r} + \frac{r^3}{2} \right] . 
\label{eq:temp} 
\eeq 
Since $\gamma$ and $E$ are constants, the temperature behaves as  
a function only of $r$ and does not depend on the shock position.
% In this way, the NS surface temperature is estimated as $ \sim 10^{9} $ K 
% for the lowest case (standard disc with critical $\dot{m}$).
% This temperature is almost consistent with former works of the accreting NS 
% surface {\bf{(need references)}}. 
% On the other hand, the inner boundary temperature becomes $ \sim 10^{11} $ K 
% for the highest case (free-fall), and it is too high to achieve in actual 
% X-ray pulsars.
% Before getting such a high temperature, the flow will emit its thermal
% energy in forms of X-ray and gamma-ray, hence in such cases the radiation 
% energy loss should be taken into account {\bf{(need references)}}.
   
We note that the shock is not in the highest-temperature part 
of the flow (although the temperatures beyond the shock are considerably 
higher than those before it) and that there are parts of the shocked 
flow at relatively lower temperatures when $r_{\rm{s}}$ becomes larger.  
Radiation from that part of the flow may appear in the low energy 
components of the observed spectra of X-ray pulsars. Also, if the shock 
heating occurs away from the stellar rotational axis, relatively hot 
fluid will exist in a broad region covering the stellar pole.  
The distribution of hot matter may change the pulse shapes: if the 
accretion flow emits radiation from a broad region, the geometry of 
the emitting flow will be changed and this may change the shape of the 
pulses (see, for example, Ref.~\citen{n81}). As noted above, the estimates for 
the  
shock location given here are not precise but a comparison between  
these shock locations and the observed properties of X-ray pulsars 
would be interesting.
 
Fig.~\ref{fig:temp} shows temperatures measured in units of $T_d$ and for 
converting these into actual temperature estimates, one needs to use suitable 
models for the disc and for the departure of the funnel flow from it. We feel 
that it is necessary to be cautious about interpreting these results directly 
in terms of actual temperatures in the flow and at the neutron star surface, 
both because of the diversity of the models and also because of the simplified 
treatment in this sub-section. However, we note that surface temperatures 
derived from this, for the models which we have been considering, lie in the 
range from $ \sim 10^{9} $ K (standard disc case; this value is consistent 
with values often quoted: 
see, for example, Ref.~\citen{bb98}) to $ \sim 10^{11} $ K (free-fall case; 
this value is too high 
for actual X-ray pulsars and would indicate that energy losses from the funnel 
flow needed to be taken into account).

% * * * * * * * * * * * * * * * * * * * * * * * * * * * *
\section{Summary and further discussion}\label{sec:7}
% * * * * * * * * * * * * * * * * * * * * * * * * * * * *

In this study, we have considered funnel-flow accretion onto a magnetized 
neutron star (NS), including taking into account the possible occurrence of a 
standing shock. We have assumed that the funnel flow goes strictly along the 
magnetic field lines and that the field is a dipole field, aligned with the 
stellar rotation axis, throughout the region of our calculation. The 
funnel flow leaves from the plane of the disc and accelerates smoothly into the 
supersonic regime, passing through a critical point. Then, at some point after 
this, a standing shock occurs beyond which the flow then accretes sub-sonically 
onto the neutron star. We have calculated the location of the standing shock, 
neglecting energy losses, and have obtained a full solution for given boundary 
conditions at the stellar surface and at the point where the funnel flow leaves 
the disc. The results suggest that the shock position depends strongly on the 
boundary conditions. As the ratio between the pressures at the inner and outer 
boundaries becomes larger, the shock location goes further from the NS surface. 
If the NS boundary pressure is high and free-fall type accretion occurs, the 
shock location must then be rather far above the stellar surface and the 
situation is as discussed by Basko and Sunyaev.\cite{bs76} In contrast, the 
column solution given by Inoue (1975) indicates that the shock would be much 
closer to the NS surface.\cite{i75} As a limiting case, Braun and Yahel 
suggested that the shock occurs just {\em on} the NS surface, and in some cases 
cannot occur at all~\cite{by84}. However, if the NS boundary pressure becomes 
quite high~\cite{bb98}, the shock front may be further out and determining its 
location will depend on complicated physics. A lot of studies of accretion flow 
onto magnetized NSs have considered flow only just near to the NS 
surface~\cite{k84}; global treatment of the flow going from the disc to the NS 
has been considered by only a limited number of 
authors.~\cite{l96,pc96,lw99,klur02}

The shock location may affect the spectra, luminosities and light curves 
of X-ray pulsars. The result that the shock location may be rather 
distant from the stellar surface has two implications for observations 
of X-ray pulsars. Firstly, as discussed in the previous section, the 
lower limit for the temperature of the shocked flow (at low temperatures 
for shocked material but much hotter than the unshocked material)  
depends only on the shock location. %% (see Fig.~\ref{fig:temp}).  
The region which has reached high temperatures due to shock heating,  
but is still optically thin, corresponds to the transition region  
discussed by Inoue and Basko \& Sunyaev.\cite{i75,bs76}. 
The spectrum of an X-ray pulsar will correspond to the  
radiation coming from the NS surface, from the optically thick 
radiative region, and from this transition region. 
The information obtained about the transition region is therefore  
relevant for modelling the observed properties of X-ray pulsars.  
Secondly, if the shock is far above the stellar surface, the 
radiating area presented by the shocked material will be large and the 
amount of radiation coming from the sides of the funnel flow will be 
significant. This radiation from the side-faces (``fan-beams'') has a 
different energy range and pulse phase compared with radiation emitted 
in the direction along the field lines (``pencil-beams''). 
This difference between the pulse phases makes the observed light curve 
complicated: having a large radiative region due to a distant shock 
location makes the fan-beam component larger, leading to irregular light 
curves with phase-inversion, twin-peaks, peak splitting and so 
on.\cite{wsh83,ww81,n81} From observed light curves and spectra, it should 
become possible to deduce the shock locations.

The analysis carried out in this paper has been very simplified. 
In actual bright X-ray sources, the complicated features which we have 
neglected may be important, especially near to the stellar surface. 
The main additional features concerned are: 
(i) {\it Radiation pressure:}~the radiation energy density can become 
quite high near to the stellar surface and radiation pressure can 
dominate over gas pressure there. 
In Section \ref{sec:add} we have estimated the effects of this  
in just a rough way; we will return to give a more thorough treatment of it in 
future work. 
(ii) {\it Radiative energy losses:}~in this paper we have neglected 
energy losses due to radiation emitted from the flow. However, this 
simplification is not valid near to the stellar surface where emissivity 
due to thermal bremsstrahlung becomes significant compared with the 
internal energy density of the flow and then, in the optically thick  
regime, black body radiation will become more efficient. These effects  
should be included in a more complete treatment. 
The energy losses may make the shock position move nearer to the star  
because they lower the pressure of the column mainly   
after the shock and hence the shock would occur further downstream. 
(iii) {\it Oblique dipole field:}~in order for a neutron star 
to be an X-ray pulsar, the dipole axis must not, in fact, coincide  
with the rotation axis. 
(iv) {\it Relativistic effects:}~neutron stars are compact objects 
with strong gravitational fields and also the funnel-flow velocity before the 
shock is quite high in some cases. Therefore both general and special 
relativistic effects may change the results quantitatively. 
(v) {\it Boundary conditions:}~there are serious uncertainties concerning the 
boundary conditions both at the stellar surface and at the point where the 
funnel flow leaves the accretion disc. 
All of the above require further work but we believe 
that the present calculations represent a useful first step and grasp 
the overall nature of these accretion flows.

\section*{Acknowledgements}

We thank S. Konar, N. Kawakatu, M. Takahashi, R. Takahashi and W. Kluzniak for 
helpful discussions. 

%\appendix
%\section{First Appendix} %Empty argument \section{} yields `Appendix'. 
%
%\section{Second Appendix}


\begin{thebibliography}{99}
%%%%%%%%%%%%%%%%%%%%%%%%%%%%%%%%%%%%%%%%%%%%%%%%%%%%%%%%%%%%%
% Some macros are available for the bibliography:
%  o for general use
%    \JL : general journals                 \andvol : Vol (Year) Page
%  o for individual journal 
%    \AJ   : Astrophys. J.           \NC         : Nuovo Cim.
%    \ANN  : Ann. of Phys.           \NPA, \NPB  : Nucl. Phys. [A,B]
%    \CMP  : Commun. Math. Phys.     \PLA, \PLB  : Phys. Lett. [A,B]
%    \IJMP : Int. J. Mod. Phys.      \PRA - \PRE : Phys. Rev. [A-E]     
%    \JHEP : J. High Energy Phys.    \PRL        : Phys. Rev. Lett.
%    \JMP  : J. Math. Phys.          \PRP        : Phys. Rep.
%    \JP   : J. of Phys.             \PTP        : Prog. Theor. Phys.     
%    \JPSJ : J. Phys. Soc. Jpn.      \PTPS       : Prog. Theor. Phys. Suppl.
% Usage:
%  \PRD{45,1990,345}          ==> Phys.~Rev.\ \textbf{D45} (1990), 345
%  \JL{Nature,418,2002,123}   ==> Nature \textbf{418} (2002), 123
%  \andvol{B123,1995,1020}    ==> \textbf{B123} (1995), 1020
%%%%%%%%%%%%%%%%%%%%%%%%%%%%%%%%%%%%%%%%%%%%%%%%%%%%%%%%%%%%%
 
\bibitem{g71}R.~Giacconi, H.~Gursky, E.~Kellogg, E.~Schreier 
and H.~Tananbaum,  \JL{Astrophys. J. Lett.,167,1971,L67}.
\bibitem{t72}H.~Tananbaum, H.~Gursky, E.~M.~Kellogg,  R.~Levinson, 
E.~Schreier and R.~Giacconi, \JL{Astropys. J. Lett.,174,1972,L143}.
\bibitem{lpp73}F.~K.~Lamb, C.~J.~Pethick and D.~Pines, 
\AJ{184,1973,271}.
\bibitem{pr72}J.~E.~Pringle and M.~Rees, \JL{A \& A,21,1972,1} .
\bibitem{wsh83}N.~E.~White, J.~H.~Swank and S.~S.~Holt, \AJ{270,1983,711}.
\bibitem{b97}L.~Bildsten, D.~Chakrabarty, J.~Chiu, M.~H.~Finger, 
D.~T.~Koh, R.~W.~Nelson, T.~A.~Prince, B.~C.~Rubin, D.~M.~Scott, 
M.~Stollberg, B.~A.~Vaughan, C.~A.~Wilson, and 
R.~B.~Wilson, \JL{Astrophys. J. Sapl.,113,1997,367}.
\bibitem{n89}F.~Nagase, \JL{PASJ,41,1989,1}.
\bibitem{st83}S.~L.~Shapiro and S.~Teukolsky, 
Black Holes, White Dwarfs and Neutron Stars, Wiley, (1983) New York.
\bibitem{bs76}M.~M.~Basko and R.~A.~Sunyaev, \JL{MNRAS,175,1976,395}.
\bibitem{i75}H.~Inoue, \JL{PASJ,27,1975,311}.
\bibitem{gl78}P.~Ghosh and F.~K.~Lamb, \AJ{223,1978,L83}.
\bibitem{lr82}S.~H.~Langer and S.~Rappaport, \AJ{257,1982,733}.
\bibitem{klur02}A.~V.~Koldoba, R.~V.~E.~Lovelace, G.~V.~Ustyugova 
and M.~M.~Romanova, \JL{Astron. J.,123,2002,2019} (KLUR).
\bibitem{t77}A.~I.~Tsygan, \JL{A \& A,60,1977,39}.
\bibitem{gpl77}P.~Ghosh, C.~J.~Pethick and F.~K.~Lamb,  
\AJ{217,1977,578}.
\bibitem{n81}W.~Nagel, \AJ{251,1981,278}.
\bibitem{ww81}Y.~M.~Wang and G.~L.~Welter, \JL{A \& A,102,1981,97}.
\bibitem{l96}J.~Li, \AJ{456,1996,696}.
\bibitem{lw99}J.~Li and G.~Wilson, \AJ{527,1999,910}.
\bibitem{kr07}W.~Kluzniak and S.~Rappaport, arXiv:0709.2361 (2007).
\bibitem{pc96}G.~Paatz and M.~Camenzind,  \JL{A \& A,308,1996,77}.
\bibitem{b52}H.~Bondi, \JL{MNRAS,112,1952,195}.
\bibitem{el77}R.~F.~Elsner, and F.~K.~Lamb, \AJ{215,1977,897}.
\bibitem{lmms86}R.~V.~E.~Lovelace, C.~Mehanian, C.~M.~ Mobarry 
and M.~E.~Sulkamen,  \JL{Astrophys. J. Supl.,62,1986,1}.
\bibitem{ll59} L.~D.~Landau and E.~M.~Lifshitz, Fluid Mechanics, 
Pergamon Press, (1959) Oxford.
\bibitem{v88}K.~A.~Van Riper, \AJ{329,1988,339}.
\bibitem{h84} A.~K.~Harding, P.~M\'{e}sz\'{a}ros, J.~G.~Kirk 
and D.~J.~Galloway, \AJ{278,1984,369}.
\bibitem{ss73}N.~I.~Shakura and R.~A.~Sunyaev, \JL{Astron. 
Astrophys.,24,1973,337}.
\bibitem{bb98} E.~F.~Brown and L.~Bildsten, \AJ{496,1998,915}.
\bibitem{gl79} P.~Ghosh and F.~K.~Lamb, \AJ{234,1979,296}.
\bibitem{w95} Y.~M.~Wang, \AJ{449,1995,L153}.
\bibitem{w87} Y.~M.~Wang, \JL{Astron. Astrophys.,183,1987,257}.
\bibitem{w96} Y.~M.~Wang, \AJ{465,1996,L111}.
\bibitem{rukl02} M.~M.~Romanova, G.~V.~Ustyugova, A.~V.~Koldoba 
and R.~V.~E.~Lovelace, \AJ{578,2002,420}.
\bibitem{by84} A.~Braun and R.~Z.~Yahel, \AJ{278,1984,349}.
\bibitem{k84} J.~G.~Kirk, \JL{Proc. ASA, 5,1984,446}.

\end{thebibliography}
\end{document}